\documentstyle[11pt,dunk2001_asp,twoside,psfig]{article}
\begin{document}
\title{Star Streams in the Milky Way Halo}
 \author{Heather Morrison}
\affil{Dept of Astronomy, Case Western Reserve University\\
10900 Euclid Ave, Cleveland OH 44118 USA}
\author{Mario Mateo}
\affil{Department of Astronomy, University of Michigan, 821
Dennison Bldg., Ann Arbor, MI 48109--1090\\
electronic mail: mateo@astro.lsa.umich.edu} 
\author{Edward W. Olszewski}
\affil{Steward Observatory, University of Arizona, Tucson,
AZ 85721\\
electronic mail: edo@as.arizona.edu} 
\author{R.C. Dohm-Palmer}
\affil{Department of Astronomy, University of Michigan, 821
Dennison Bldg., Ann Arbor, MI 48109--1090\\
electronic mail: rdpalmer@astro.lsa.umich.edu}
\author{Paul Harding}
\affil{Steward Observatory, University of Arizona, Tucson, Arizona 85726
\\ electronic mail: harding@billabong.astr.cwru.edu}
\author{Amina Helmi}
\affil{Max Planck Institut fur Astrophysic, Garching bei Munchen,
Germany\\
electronic mail: ahelmi@mpi-garching.mpg.de}
\author{John Norris}
\affil{Research School of Astronomy and Astrophysics, ANU, Private Bag,
Weston Creek PO, 2611 Canberra, ACT, Australia\\
electronic mail: jen@mso.anu.edu.au}
\author{Kenneth C. Freeman}
\affil{Research School of Astronomy and Astrophysics, ANU, Private Bag,
Weston Creek PO, 2611 Canberra, ACT, Australia\\
electronic mail: kcf@mso.anu.edu.au}
\author{Stephen A. Shectman}
\affil{Carnegie Observatories, 813 Santa Barbara St, Pasadena, CA,
91101\\
electronic mail: shec@ociw.edu}

\newpage

\begin{abstract}

The last 10-20 years has seen a profound shift in views of how the
Galaxy's halo formed.  The idea of a monolithic early collapse of a
single system (Eggen, Lynden-Bell and Sandage 1962) has been
challenged by observations at high redshift and by cosmological models
of structure formation.  These findings imply that we should see clear
evidence of hierarchical formation processes in nearby galaxies.
Recent studies of our Galaxy, made possible by large-scale CCD surveys
such as the Sloan Digital Sky Survey (SDSS), have begun to reveal
tantalizing evidence of substructure in the outer halo.  We review
evidence for tidal streams associated with known Milky Way satellites
and for star streams whose progenitors are still unknown. This 
includes results from the SDSS and our own ongoing pencil-beam halo
survey, the Spaghetti survey.

\end{abstract}

\section{Introduction}

A revolution is underway in our understanding of how galaxies form:
The idea of a monolithic collapse of a single system at earliest times
(Eggen et al. 1962) has been challenged by observations
at high redshift and by increasingly sophisticated models of the
evolution of structures within the basic framework of the Hot Big Bang
model (for example, Pearce et
al. 1999, Steinmetz \& Navarro 1999).  These findings imply that we
should see clear evidence of hierarchical formation processes within
nearby galaxies.  Indeed, some of the earliest work that suggested
complex and possibly hierarchical evolutionary histories for galaxies
such as ours came from studies of the stellar populations of the outer
halo of the Milky Way (Searle and Zinn 1978).  Studies
of nearby and high-redshift systems represent complementary
approaches.

We can learn much about accretion processes in galaxy formation using
the Milky Way because we can view our halo in 3 dimensions. By adding
the kinematics of halo stars to their 3-D positions, moving groups are
given enhanced contrast (see, for example, Harding et al. 2001).  This
sort of study is impossible for any other spiral galaxy.

Recent studies of our Galaxy have begun to reveal tantalizing evidence
of possible substructure in the halo, both locally (Majewski 1992, C{\^o}t{\'e} et al. 1993, Arnold and Gilmore 1992, Helmi
et al. 1999) and at larger distances
(Ivezi{\' c} et al. 2000, Yanny et al. 2000).
And of course, the Sgr dwarf galaxy (Ibata et al. 1995) is 
injecting globulars and field stars
that formed in a completely separate entity from the Milky Way into
its halo.  The question has now shifted from `Did accretion play any
role in the formation of the Galaxy?'  to `How much of the Galaxy
formed from accretion of hierarchical fragments?'  We will show below
that the data that is needed to answer the second question
definitively also has the potential to teach us about the dark matter
halo of the Galaxy and may directly constrain cosmological formation
models such as that of Klypin et al. (1999), Moore et al. (1999).
 
\section{The ``Spaghetti'' Survey}
 
Several years ago we began a search designed to find halo streams if
they are present.  Since then, the Sloan Digital Sky Survey (SDSS)
team has found one stream in their first few hundred degrees of
equatorial test data (Ivezi{\' c} et
al. 2000, Yanny et al. 2000).  Our group has found
evidence that has strengthened the case that this substructure is 
associated with the Sgr dwarf using velocities of  halo
giants (Dohm-Palmer et al.2001).  Our survey is unique in using a set of
halo tracers (giants) that can be observed spectroscopically out 
to  distances of hundreds of kpc where it is likely that only dwarf
galaxies are found.

Our photometric survey consists of several hundred pencil beams at
high Galactic latitude, which will comprise 100 square degrees of data
in the Washington system.  Our goal is to use
this survey, the spectroscopic followup, and our
extensive modelling, to quantify the amount of halo substructure
associated with completely destroyed satellites, with partially
destroyed satellites like the Sgr dwarf, and with intact objects such
as $\omega$ Cen. More details of the survey are available in
Morrison et al. (2000), Dohm-Palmer et
al.(2000), Morrison et al. (2001) and at
http://www.astro.lsa.umich.edu/users/rdpalmer/spag/. In this paper we
focus on the part of the survey which identifies distant halo giants.

Local studies of halo stars passing through the solar neighborhood are
limited in numbers of stars, do not well sample all regions of the
halo, and often have biased kinematics.  Chiba and Yoshii(1998) have
shown that the 250 red giants and RR Lyrae stars with space motions
determined from Hipparcos occupy a region extending to at most 35 kpc
from the Galactic center.  Carney et al. (1996) derive the
apogalacticon distances of approximately 1200 local stars.  There are
only 23 stars in this sample with calculated apogalacticon distances
larger than 30 kpc.  These particular stars, by definition, are on
extremely elongated orbits.  There are vast parts of the Milky Way,
and vast parts of phase space, therefore left unsampled.  Figure 1
illustrates this clearly: halo globular clusters are
plotted with open symbols, and it can be seen that they extend out to
almost 100 kpc from the Galactic center. Known field stars 
(taken from the compilation of Beers \& Sommer-Larsen 1995) are shown
with closed circles. Most halo field stars known are within a few kpc
of the Sun.

\begin{figure}
\centerline{\vbox{
\psfig{figure=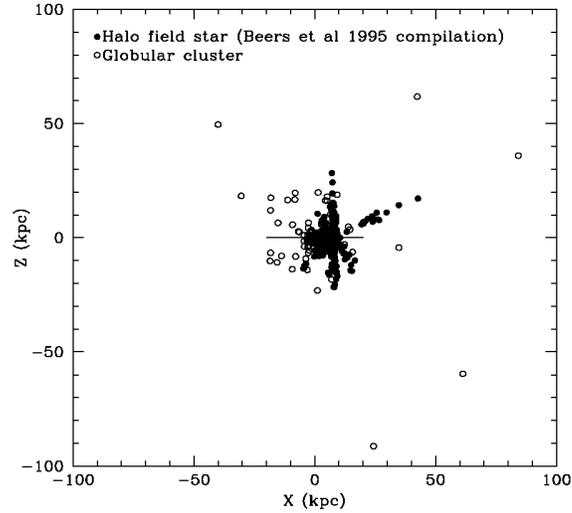,width=3in}
}}
        
\caption{Halo field stars known in 1995, from the compilation of Beers
\& Sommer-Larsen (1995), plotted with closed circles. Comparison with
the distribution of globular clusters (open circles) shows that even
though halo field stars are likely to stretch out to $\sim$100 kpc
from the Sun, very few are known in the outer halo.
\label{halo95}}
\end{figure}

\begin{figure}
\centerline{\vbox{
\psfig{figure=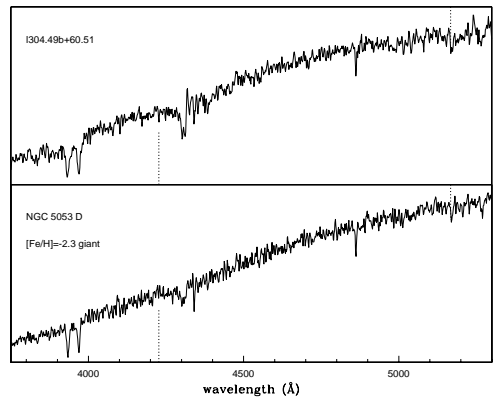,width=3in,angle=270}
}}
        
\caption{Spectrum of one of the most distant halo giants we have
identified, star l304.49b+60.51, which has a metallicity of
[Fe/H]$\simeq$--1.9 and a distance of $\sim$80 kpc from the Sun. Shown
for comparison is the spectrum of a giant of similar color in the most
metal-poor globular cluster known, NGC 5053 D. The features that we
use for luminosity discrimination, CaI 4227 and the Mg$b$/MgH feature,
are marked.
\label{800}}
\end{figure}
 
Figure 2 shows the spectrum of one of the most distant giants
we have identified so far, star l304.49b+60.51, which lies at a distance of
$\sim$80 kpc from the Sun.
We expect to identify several hundred giants from the extreme outer halo when
our 100 square degree survey is complete: an order of magnitude more
than are currently known.

\section{Searching for Substructure in the Outer Halo}

Halo red giants are useful probes because of their high luminosity,
which makes spectroscopic followup much easier. However, they are
rare: we average $\sim$1 giant in each 0.25 square degree field. This
means that we cannot use many of the techniques which were developed to search
for substructure in the larger local samples. These techniques are
often one- and two-dimensional variations on the use of velocity
histograms, whether to search for a spike (C{\^o}t{\'e} et al. 1993), unusual
outliers (Majewski 1992) or simply a non-gaussian velocity distribution
(Harding et al. 2001). 

We have used the extensive simulations of Harding et al. (2001),
re-normalized so that one particle corresponds to one halo giant, to
help us identify and test techniques to search for substructure. Our
method uses the fact that that there are strong correlations between
position and velocity along a disrupting stream, illustrated in Figure
3. Our sparse sampling reduces the numbers of stars detected in a
given stream, but preserves the position/velocity correlations, as can
be seen in Figure 3.

\begin{figure}
\centerline{\vbox{
\psfig{figure=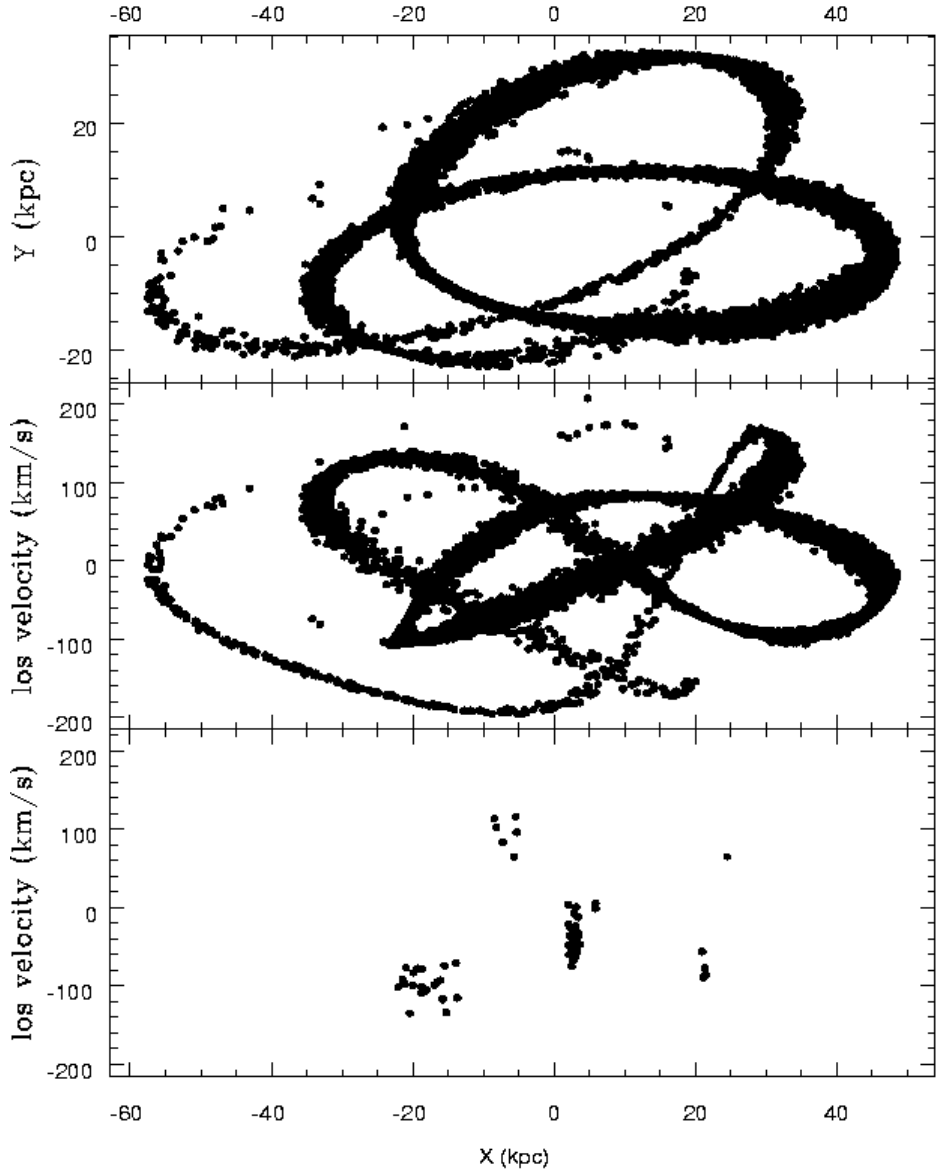,width=5in}
}}

\caption{Disruption of a stream in position and velocity space. The
top panel plots the X and Y galactic coordinates of points along
several wraps of the stream. The middle panel shows the line-of-sight
velocity plotted against X, illustrating the velocity gradients along
the stream. The bottom panel shows the particles that would be
observed in a 100 square degree survey such as ours, with realistic
errors added in both distance and velocity. Although the picture is
less clean, correlations between position and velocity still persist.
\label{stream}}
\end{figure}

Since we can measure the distances to our giants with reasonable
accuracy,  we are able to derive all three spatial
coordinates, but because they are too distant for proper motions until
space missions such as GAIA and SIM are operating, we only have one of the
three velocity coordinates, the line-of-sight radial velocity
$V_{los}$. We define a scaled metric in these four dimensions to
measure the correlations between distance and velocity: two stars with
similar distances and velocities will have a very small 4-D
``distance'' apart. 

For two stars with galactic latitude, galactic longitude, distance and
velocity $l_1,b_1,d_1,V_1$ and $l_2,b_2,d_2,V_2$, we write the metric
as:

\[ D_{lbdV} = \left( w_1(l_1-l_2)^2 +
w_2(b_1-b_2)^2 +w_3(d_1-d_2)^2 + w_4(V_1-V_2)^2 \right)^{1/2} \]

We choose the scalings $w_i$ for the l, b, distance and velocity terms to
account for the different measurement accuracy of each quantity, and
for the fact that streams can spread spatially but have a tight
relation between position and velocity at a given location. We
know l and b very accurately, velocity to $\sim$10\% (20 km/s) and
distance to $\sim$25\%, so two values that differ by less than their
errors will be given less weight than values that differ
significantly. 

We use two different simulated datasets to test this technique, with
sample sizes equal to our expected number of giants when we have
completed our survey.  The first contains 500 points from an accreted
halo simulation constructed with 23 disrupting tidal streams from the
library of Harding et al. (2001). These points were chosen from a
dataset containing 320,000 points in 23 streams in a way that
simulates our real observations.  We used our real field centers for
the 50 square degrees where observations have been taken, and planned
field centers plus reflections about the galactic plane to make 400
pencil-beam field centers at high galactic latitude. If the stream
particle was in the 0.25 square degrees surrounding a field center, we
applied observational errors of 20 km/s for velocity and 25\% for
distance. The second dataset contained points from a smooth halo
constructed with an isotropic velocity ellipsoid, as described in
Harding et al. (2001), chosen to simulate our real observations in the
same way.

We calculate the value of the scaled 4-d metric described above for
every pair of points in the smooth and accreted halo simulations. As
expected, the accreted halo shows significantly more points with
small values of the metric, as can be seen in Figure
4. We are working with Jiayang Sun (CWRU statistics) to develop
sensitive statistical tests for this accretion signature.

\begin{figure}

\centerline{\vbox{
\psfig{figure=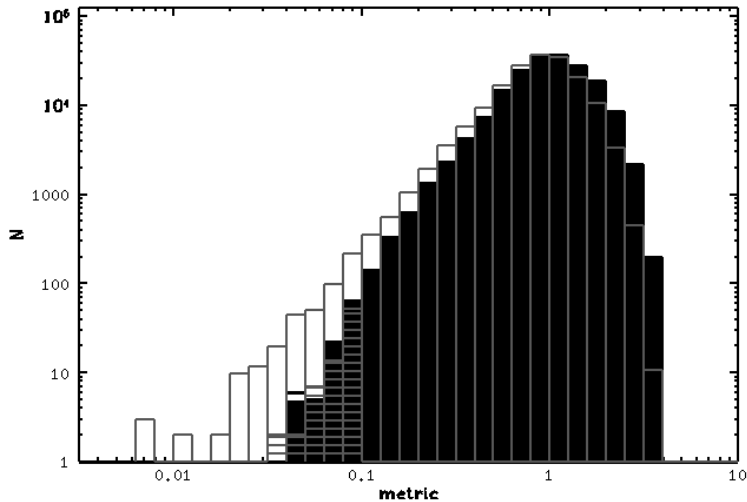,width=4in,angle=270}
}}

\caption{Comparison of 4-d metric values for a smooth halo (filled
histogram) and for an accreted halo (open histogram). (Cross-hatched
histogram shows particles matched with particles from another
stream). It can be seen
that the accreted halo has significantly more points with small 4-d
distances, as expected.
\label{loghist}}
\end{figure}

\section{First Results}

Here we show some preliminary results for our sample of 58 halo giants
with spectroscopic confirmation. The distances are calculated using
the photometric metallicities only at this point, although we will
soon have spectroscopic metallicities available, which will lead to
more accurate distance measures. Figure 5 shows the
histogram of metric values for this sample (calculated as described
above) and the histogram of metric values for a smooth halo sample
with the same size as our real halo giant sample, chosen from the
simulated smooth halo described above. While the sample size is small,
we are already seeing some of the signatures of an
accreted halo.

\begin{figure}

\centerline{\vbox{
\psfig{figure=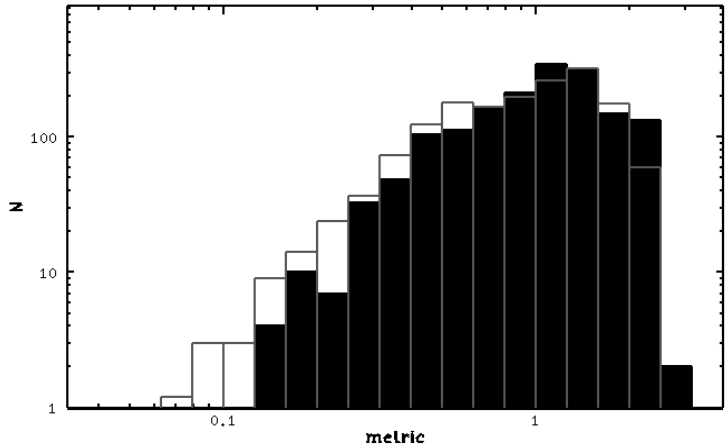,width=4in,angle=270}
}}

\caption{Comparison of 4-d metric values for a smooth halo (filled
histogram) and for our preliminary halo giant sample (open histogram).
The smooth halo was constructed to have the same sample size as our
real sample. Although the sample size is small, we are already seeing
indications of some clustering in the 4-d velocity/distance space.
\label{realhist}}
\end{figure}

Some of the smallest values of the 4-d metric in our giant sample come
from a group of stars which are located in the same region of the sky
as the SDSS overdensity (Ivezi{\' c} et al. 2000, Yanny et
al. 2000). This gives us the opportunity to test suggestions that this
stream is associated with the Sgr dwarf by comparing the velocities of
giant stars at similar distances to the SDSS BHB and RR Lyrae stars to
the predictions of models of the disruption of Sgr. We have chosen to
use the models of Helmi \& White (2001), which, unlike earlier models
of its disruption, can explain the fact that the galaxy was not
completely disrupted on its first few passages without resort to
unusual scenarios such as a very stiff and extended dark matter halo
(Ibata \& Lewis 1998) or a convenient recent collision with the Large
Magellanic Cloud (Zhao 1998).

Figure 6 shows the predictions of the purely stellar model of Helmi \&
White (2001) for particles close to the minor axis of the Galaxy,
compared with the actual positions and velocities of 21 confirmed
metal-poor giants in this direction. It can be seen that all of the
halo giants with distances greater than 30 kpc agree well with the
predictions of the Sgr disruption model, confirming the suggestions of
Ivezi{\' c} et al. (2000) and Ibata et al. (2001a) that the SDSS
stream is associated with Sgr. The right-hand panel of the plot shows,
for comparison, one simulation taken from a smooth halo: it can be seen
that the arrangement of distances and velocities that we observe is
very unlikely to have occurred by chance. More details of these
observations can be found in Dohm-Palmer et al. (2001).

The wrap at distances near 50 kpc is the one detected by the SDSS. It
is interesting to note that we may also have detected stars from the
wraps near 20 kpc (although the large velocity range predicted for
these wraps makes this harder to test) and from the wrap at 80 kpc,
which contains debris lost from the Sgr dwarf on one of its earliest
passages. Such debris, if confirmed with a larger sample, will give us
the opportunity to measure the M/L ratio of the progenitor to the Sgr
dwarf, as the outer regions are the first to be lost to tidal
disruption, and the presence of a massive dark matter halo in the
outer regions will help retain more stars than in a galaxy with little
or no dark matter (Helmi 2000).

Our sample of giants from 100 square degrees of imaging will be large
enough to assign membership to different streams in the outer
halo. This will allow their use in ways that were almost inconceivable
ten years ago. For example, we can learn about the structure of the
progenitor satellites: the red giants that are Sgr stream members have
a mean metallicity of [Fe/H]=--1.5, significantly more metal-poor than
the main body of Sgr (Layden and Sarajedini 2000). This suggests that
the Sgr progenitor had a radial abundance gradient. As mentioned
above, identification of stars lost on early passages of Sgr will
allow us to constrain the mass-to-light ratio of its progenitor. The
spread in longitude of the Sgr debris will constrain the flattening of
the Galaxy's dark matter halo (Helmi 2000, Ibata et al. 2001b). We can
even use the stream velocity dispersion to constrain the lumpiness of
the Galaxy's dark halo, which will allow us to {\it directly} test the
predictions of simulations of structure formation (eg Klypin et
al. 1999, Moore et al. 1999).

\begin{figure}

\centerline{\vbox{
\psfig{figure=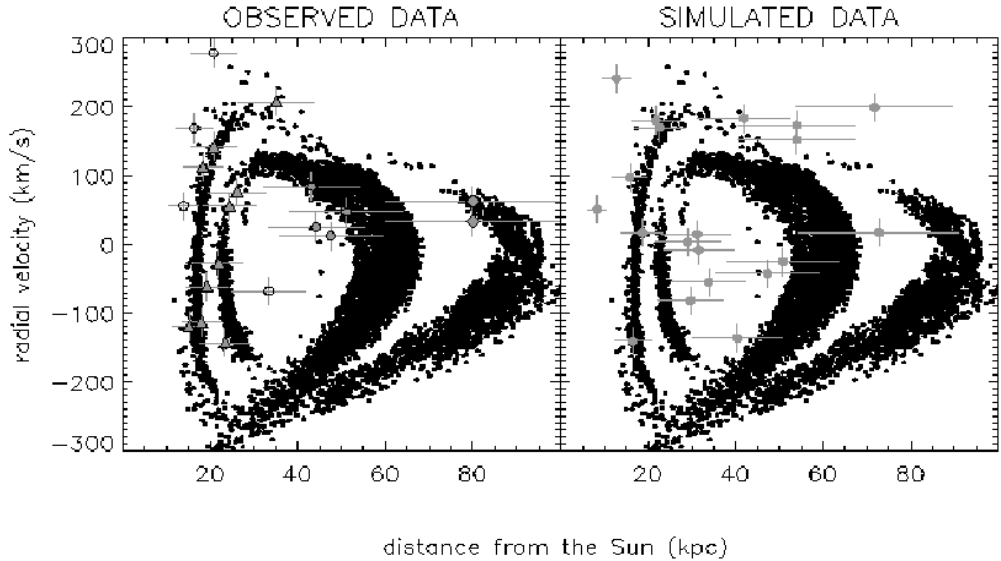,width=6in,angle=270}
}}

\caption{(Left panel) Comparison of the purely stellar model of the Sgr dwarf and
its disruption from Helmi \& White (2001) (closed black symbols) with the 21
halo giants we have found in this direction (grey symbols with error
bars). The right panel shows the comparison of the model predictions
with a typical  simulation of a smooth halo.  It is very unlikely that
our data came from a smooth halo.
\label{paper5}}
\end{figure}





\acknowledgements

The Galaxy's halo and its formation is a subject which is close to
Ken's heart: he supervised two dissertations on the topic
(Ratnatunga's and Morrison's). The technology available at the time
(photographic plates) wasn't equal to the challenge of studying the
outer halo, so it's a particular pleasure to be able to talk about a
successful survey (and one that he is involved in) for his 60th year
celebration. It looks like his early interest in accreted halos was
spot on: the instrumentation just had to catch up!

HLM, EWO and MLM acknowledge support from the NSF on grants AST
96-19490, AST 00-98435, AST 96-19524, AST- 00-98518, AST 95-28367, AST
96-19632 and AST 00-98661.


\begin{references}


\reference Arnold, R.\ and Gilmore, G.\ 1992, \mnras, 257, 225 
\reference Beers, T.~C.~\& 
Sommer-Larsen, J.\ 1995, VizieR On-line Data Catalog: J/ApJS/96/175.
Originally published in: \apjs, 96, 175 

\reference Carney, B.\ W., Laird, J.\ B., Latham, D.\ W.\ and Aguilar, L.\ A.\ 1996,
\aj, 112, 668
\reference Chiba, M.\ and
Yoshii, Y.\ 1998, \aj, 115, 168

\reference C{\^o}t{\'e}, P., Welch, D.L.,
Fischer, P. and Irwin, M.J. 1993, \apjl, 406, L59 

\reference Dohm-Palmer, R.~C., 
Mateo, M., Olszewski, E., Morrison, H., Harding, P., Freeman, K.~C., \& 
Norris, J.\ 2000, \aj, 120, 2496 
\reference Dohm-Palmer, 
R.~C.~et al.\ 2001, \apjl, 555, L37 

\reference Eggen, O. J., Lynden-Bell, D., \&
Sandage, A. R. 1962, \apj, 136, 748

\reference Harding, P., Morrison, 
H.~L., Olszewski, E.~W., Arabadjis, J., Mateo, M., Dohm-Palmer, R.~C., 
Freeman, K.~C., \& Norris, J.~E.\ 2001, \aj, 122, 1397 
\reference Helmi, A.\ 2000, doctoral dissertation,
Leiden University
\reference Helmi, A., White, S.\ D.\ M., de Zeeuw, P.\ T.\ and
Zhao, H.\ 1999, Nature, 402, 53 
\reference Helmi, A.~\& White, S.~D.~M.\ 2001, \mnras, 323, 529

 
\reference Ibata, R.\
A., Gilmore, G.\ and Irwin, M.\ J.\ 1995, \mnras, 277, 781
\reference Ibata, R.~A.~\& Lewis, 
G.~F.\ 1998, \apj, 500, 575 
\reference Ibata, 
R., Irwin, M., Lewis, G.~F., \& Stolte, A.\ 2001a, \apjl, 547, L133 
\reference Ibata, R., Lewis, G.~F., 
Irwin, M., Totten, E., \& Quinn, T.\ 2001b, \apj, 551, 294 




\reference Ivezi{\' c}, {\v Z}eljko et al.\ 2000, \aj, 120, 963 

\reference Klypin, A., Kravtsov, A.\ V., Valenzuela, O.\ 
and Prada, F.\ 1999, \apj, 522, 82 

\reference Layden, A.~C.~\& Sarajedini, A.\ 2000, \aj, 119, 1760 
 

\reference Majewski, S.\ R.\ 1992, \apjs, 78, 87
\reference Moore, B., Ghigna, S., 
Governato, F., Lake, G., Quinn, T., Stadel, J.\ and Tozzi, P.\ 1999, \apjl, 
524, L19 
\reference Morrison, H.~L., 
Mateo, M., Olszewski, E.~W., Harding, P., Dohm-Palmer, R.~C., Freeman, 
K.~C., Norris, J.~E., \& Morita, M.\ 2000, \aj, 119, 2254 


\reference Morrison, H.~L., 
Olszewski, E.~W., Mateo, M., Norris, J.~E., Harding, P., Dohm-Palmer, 
R.~C., \& Freeman, K.~C.\ 2001, \aj, 121, 283 


\reference Pearce F.R., Jenkins A., Frenk 
C.S., Colberg J.M., White S.D.M., Thomas P.A., Couchman H.M.P., 
Peacock J.A., Efstathiou G. 1999, \apjl, 521, 99
\reference Searle, L. \& Zinn, R. 1978, ApJ,
225, 357
\reference Steinmetz M. \& Navarro 
J.F. 1999, \apj, 513, 555
\reference Yanny, B., Newberg, H. et al.\ 2000, 
\apj, 540, 825 

\reference Zhao, H.\ 1998, \apjl, 500, L149 

\end{references}
\end{document}